\documentclass[twocolumn,nofootinbib,aps,prl,floats,floatfix,amsmath,amssymb,longbibliography,secnumarab
ic,superscriptaddress,preprintnumbers]{revtex4-1} %
\usepackage[final]{graphicx}
\usepackage{hyperref}
\usepackage{amsmath}
\usepackage{bbm}
\usepackage{bm}
\usepackage{amsfonts}
\usepackage{amssymb}
\usepackage{latexsym}
\usepackage{graphicx}
\usepackage[english]{babel}
\usepackage{multirow}
\usepackage{float}
\usepackage{url}
\usepackage{slashed}
\usepackage{xcolor} 
\usepackage[utf8]{inputenc}
\usepackage{stmaryrd} 
\usepackage{enumitem}
\usepackage{hyperref}
\usepackage{cleveref}
\usepackage{siunitx}
\usepackage{verbatim}

\usepackage{amsthm}

\theoremstyle{remark}

\newcommand{\be}{\begin{equation}}
\newcommand{\ee}{\end{equation}}
\newcommand{\ba}{\begin{array}}
\newcommand{\ea}{\end{array}}
\newcommand{\bea}{\begin{eqnarray}}
\newcommand{\eea}{\end{eqnarray}}

\newcommand{\besub}{\begin{subequations}}
\newcommand{\eesub}{\end{subequations}}

\definecolor{darkerblue}{rgb}{0.2,0.2,0.5}
\definecolor{seagreen}{rgb}{0.180392,0.545098,0.341176}
\definecolor{smagenta}{rgb}{0.5,0.145098,0.341176}
\definecolor{deepblue}{rgb}{0,0,1}

\begin{document}

\title{Radio-frequency Dark Photon Dark Matter across the Sun }

\author{Haipeng An}
\email{anhp@mail.tsinghua.edu.cn}
\affiliation{Department of Physics, Tsinghua University, Beijing 100084, China}
\affiliation{Center for High Energy Physics, Tsinghua University, Beijing 100084, China}

\author{Fa Peng Huang}
\email{fapeng.huang@wustl.edu}
\affiliation{Department of Physics and McDonnell Center for the Space Sciences, Washington University, St. Louis, MO 63130, USA}
\affiliation{TianQin Research Center for Gravitational Physics and School of Physics and Astronomy, Sun Yat-sen University (Zhuhai Campus), Zhuhai 519082, China}

\author{Jia Liu}
\email{jialiu@pku.edu.cn}
\affiliation{School of Physics and State Key Laboratory of Nuclear Physics and Technology, Peking University, Beijing 100871, China}
\affiliation{Center for High Energy Physics, Peking University, Beijing 100871, China}

\author{Wei Xue}
\email{weixue@ufl.edu}
\affiliation{Department of Physics, University of Florida, Gainesville, FL 32611, USA}

\begin{abstract}

Dark photon as an ultralight dark matter candidate can interact with the Standard Model particles via kinetic mixing. 
We propose to search for the ultralight dark photon dark matter using radio telescopes with solar observations.
The dark photon dark matter can efficiently convert into photons in the outermost region of the solar atmosphere, the 
solar corona, where the plasma mass of photons is close to the dark photon rest mass. 
Due to the strong resonant conversion and benefiting from the short distance between the Sun and the Earth, 
the radio telescopes can lead the dark photon search sensitivity in the mass range of $4 \times 10^{-8} -  4\times 10^{-6} \, \rm{eV}$, 
corresponding to the frequency $10 - 1000 \, {\rm MHz}$. As a promising example, the operating radio telescope LOFAR 
can  reach the kinetic mixing $\epsilon \sim  10^{-13}$  ($10^{-14}$) within 1 (100) hour solar observations.
The future experiment SKA phase 1 can reach $\epsilon \sim  10^{-16} - 10^{-14}$ with $1$ hour solar observations.

\end{abstract}
\maketitle

\noindent \textit{\textbf{Introduction}}-- The ultralight bosonic fields are attractive dark matter (DM) candidates.
Within them, the QCD axions, axion-like particles, and dark photons are well-studied scenarios~\cite{Essig:2013lka, Battaglieri:2017aum}.  
Kinetic mixing dark photon is one of the simplest extension of new physics beyond the Standard Model (SM) 
via a marginal operator, which is well-motivated at low energies. It can also constitute DM~\cite{Holdom:1985ag, Redondo:2008ec, Nelson:2011sf, Arias:2012az, Graham:2015rva}
and may reveal the theories beyond the SM~\cite{Dienes:1996zr, Abel:2003ue, Abel:2006qt, Abel:2008ai, Goodsell:2009xc}. There are many searches looking for dark photon or dark photon DM.
For mass $\lesssim 10^{-9}$ eV, the dark photon DM can be constrained by the observation of astronomical radio sources~\cite{Lobanov:2012pt}, CMB spectrum distortion, BBN, Lyman-$\alpha$ and heating of primordial plasma~\cite{Mirizzi:2009iz, Arias:2012az, Kunze:2015noa, Dubovsky:2015cca, Kovetz:2018zes, Pospelov:2018kdh, McDermott:2019lch, Caputo:2020bdy, Garcia:2020qrp}.
In the optical mass range of $0.1-10~{\rm  eV}$, dark photon DM can be detected by the optical haloscope~\cite{Baryakhtar:2018doz}. For dark photon DM with a mass larger than about ${\cal O}(10)$ eV, it can be absorbed in the underground DM detectors and produce electronic recoil signals~\cite{Pospelov:2008jk,An:2014twa,Bloch:2016sjj,Aprile:2019xxb}.
Dark photon lighter than the temperatures at the center of stars can also be produced inside stars and suffer stellar cooling constraints~\cite{An:2013yfc,Redondo:2013lna,Vinyoles:2015aba,An:2020bxd}.
Dark photons produced inside the Sun can be detected by DM direct detection experiments~\cite{An:2013yua, She:2019skm, An:2020bxd}.

In this letter, we focus on the radio mass window ($ 10^{-8} - 10^{-6}~ {\rm eV}$) for dark photon and assume it constitutes all the DM. 
This mass window is of particular interest because it overlaps with the regions that dark photon DM is naturally 
produced by mechanisms including the inflationary fluctuations~\cite{Graham:2015rva, Ema:2019yrd}, parametric resonances~\cite{Co:2018lka, Dror:2018pdh, Bastero-Gil:2018uel, Agrawal:2018vin},  
cosmic strings~\cite{Long:2019lwl}, the misalignment with non-minimal coupling to the gravity~\cite{Arias:2012az, AlonsoAlvarez:2019cgw} (see the ghost instability discussion in \cite{Nakayama:2019rhg}), and 
production by inflaton motion~\cite{Nakai:2020cfw}. 
The relevant searches for dark photon DM are haloscope experiments~\cite{DePanfilis:1987dk, Wuensch:1989sa, Hagmann:1990tj, Asztalos:2001tf, Asztalos:2009yp, Nguyen:2019xuh}, dish antenna experiments~\cite{Horns:2012jf, Knirck:2018ojz}, plasma telescopes~\cite{Gelmini:2020kcu} and CMB spectrum distortion~\cite{Arias:2012az, McDermott:2019lch}. 
The searches include direct detection of local dark photon DM in laboratories and observation on its impact in the early universe. 
Differently, we proposal to look for resonant conversion of dark photon DM $A' \to \gamma$ at the Sun through the radio telescopes for solar observations. This is an indirect detection of dark photon DM signal from the closest astronomical object, the Sun. It provides competitive sensitivities even with existing radio telescopes and opens vast new parameter space with future setups.

\begin{figure}
	\centering
	\includegraphics[width=1. \columnwidth]{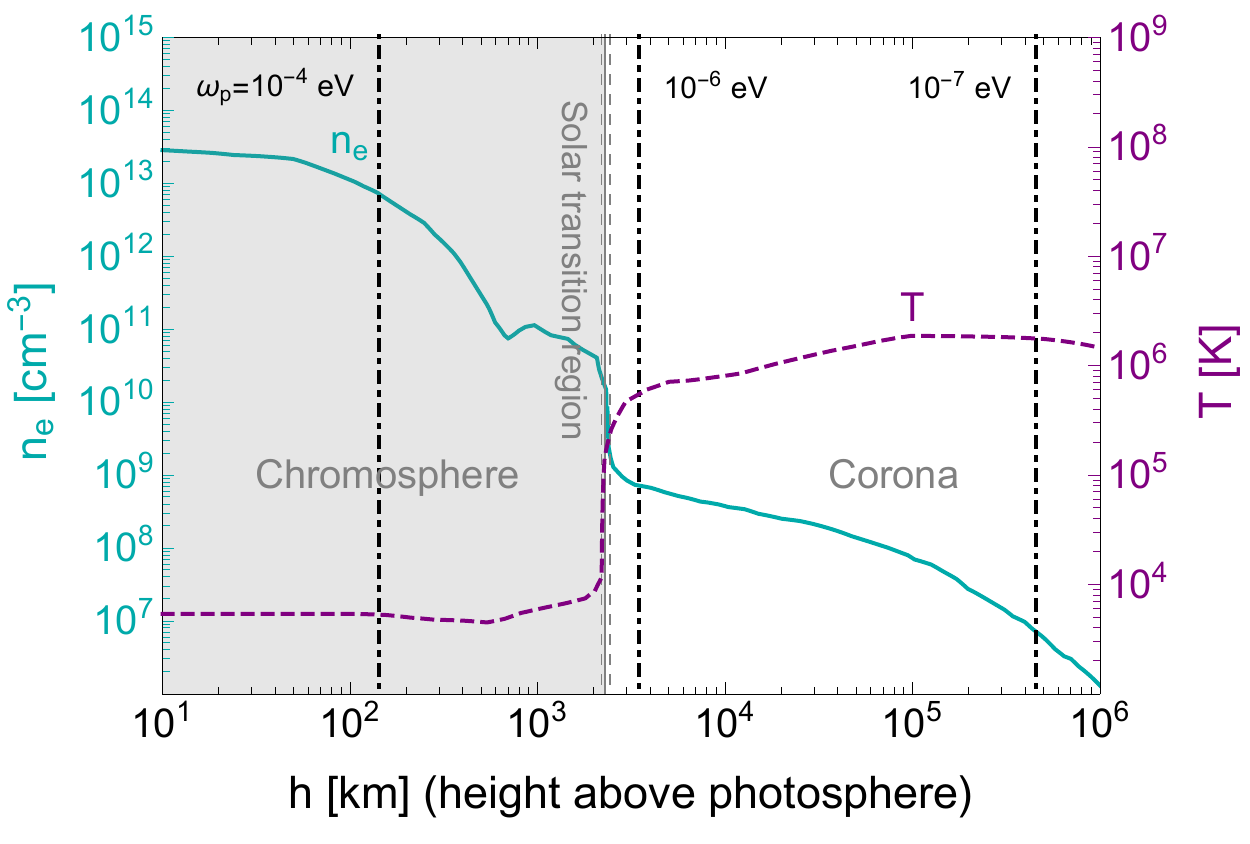} 
	\caption{The electron number density (green solid line)  and the temperature (purple dashed line) distribution for the quiet Sun from \cite{2008GeofI..47..197D}. In the gray shaded region, the converted photons produced below the solar transition region cannot propagate out of the Sun. 
	The radius for photon plasma mass $\omega_p = 10^{-4} \, , 10^{-6}\, , 10^{-7}\, {\rm eV}$
      are shown in vertical dot-dashed lines. The height above the photosphere $h$ and
      the radius $r$ has the relation $r \equiv h + r_{\rm ps}$, where $r_{\rm ps} = 695,510 \, {\rm km}$ is the radius for the solar photosphere.  
	}	\label{fig:ne}
\end{figure}

Below the electroweak scale, the minimal coupling between the  dark photon and the Standard Model particles can be described by the following Lagrangian density
\begin{equation}
   \mathcal{L} =  - \frac{1}{4} F'_{\mu\nu} F'^{\mu\nu}  - \frac{1}{2} m_{A'}^2 A'_{\mu} A'^{\mu} -\frac{1}{2} \epsilon  F_{\mu \nu} F'^{\mu \nu} \ ,
\end{equation}
where $F_{\mu \nu}$ is the photon field strength, $A'$ is the dark photon field, $F'^{\mu \nu}$ is the dark photon field strength, and $\epsilon$ is the kinetic mixing. With this mixing term, the dark photons can oscillate resonantly into photons in thermal plasma once the plasma frequency $\omega_p \approx m_{A'}$. The plasma frequency for non-relativistic plasma relies on the electron density $n_e$, 
\begin{align}
\omega_p = \left(\frac{4\pi \alpha n_e}{m_e} \right)^{1/2} 
      = \left( \frac{ n_e} { 7.3 \times 10^{8}  \, {\rm cm}^{-3}} \right)^{1/2}
\, \mu \mathrm{eV}  \ ,
\label{eq:photonmass}
\end{align}
where $\alpha$ and $m_e$ are the fine structure constant and electron mass, respectively. In the Sun's corona, $n_e\sim10^6-10^{10}$ cm$^{-3}$ is shown in Fig.~\ref{fig:ne}. Hence the range of the plasma frequency $\omega_p$ is 
from $4\times10^{-8}$ to $4\times 10^{-6}~{\rm eV}$. If $m_{A'}$ falls in this range, $A'$ can resonantly convert into a monochromatic radio wave in the corona, with the peak frequency corresponding  to $m_{A'}$, which is in the range of about $10-1000~{\rm MHz}$. This frequency 
range happens to be in the sensitive region of the terrestrial radio telescopes, such as the LOw-Frequency ARray (LOFAR)~\cite{vanHaarlem:2013dsa} and Square Kilometer Array (SKA)~\cite{SKA1-tech}. Therefore, we propose to use radio telescopes to search for dark photon DM in this mass range.
\\

\noindent \textit{\textbf{Resonant Conversion in the Sun's Corona}}--
The average conversion probability of a dark photon particle flying across the Sun's corona is the time integral of the decay rate
of $A' \to \gamma$, written as
\begin{eqnarray}
   \label{eq:Prate}
   P_{A' \to \gamma} ( v_r )
      &=& \int \frac { {\rm d} t }{2 \omega} \frac{ {\rm d}^3 p }{ (2 \pi )^3 2 \omega } 
       ( 2\pi)^4 \delta^{4} \left( p_{A'}^\mu - p_\gamma^\mu \right) \,  \frac{1}{3}{\sum_{\rm pol}}|{ \cal M} |^2  
      \nonumber  \\
      &=& \frac{2}{3} \times 
      { \pi \, \epsilon^2 \, m_{A'} }
      \, v_r^{-1} \,
         \left| \frac{\partial \ln \omega_p^2 ( r )}  
   {   \partial r     }  \right|^{-1}_{\omega_p(r) = m_{A'}} \ .
\end{eqnarray}
Here we take average of the initial state of $A'$. During the structure formation, 
the momentum direction of $A'$ is randomly rotated in the gravitational potential. Therefore, each mode (either transverse or longitudinal) has the equal probability, $1/3$. Since only the transverse modes of photon can survive outside the plasma and propagate to the Earth, we only sum over the transverse polarizations in the final state. 
In the second line, $v_r$ is the velocity projected on the radial direction of the Sun. 
Due to the spherical distribution of $n_e$, $\omega_p$ only changes in the radial direction.

In Eq.~\eqref{eq:Prate}, it utilized the quantum field method to calculate
the $1 \to 1$ conversion and the matrix element $\cal{M}$ is derived by directly using the kinetic mixing 
operator $\frac{1}{2}\epsilon F_{\mu \nu}' \, F^{\mu \nu}$.
Due to the momentum conservation, it only applies for the resonant conversion $\omega_p = m_{A'}$.
An equivalent way to calculate the conversion rate is to solve the linearized wave equations for the photon and 
dark photon \cite{Raffelt:1987im}, which works for both resonant and non-resonant conversion.  
After applying the saddle point approximation, the result is the same as in Eq.~\eqref{eq:Prate}.
It can be explicitly shown that the non-resonant contribution is negligible.
The detailed calculations for the two methods are given in the \textit{Supplemental Material}. 
Finally, the above result is in agreement with the probability for inverse conversion $\gamma \to A'$~\cite{Mirizzi:2009iz}.

Given the conversion probability, the radiation power $\mathcal{P}$ per
solid angle $d\Omega$ at the conversion radius $r_c$ is 
\begin{align}
   \frac{d \mathcal{P}}{d \Omega} &\approx
         2 \times 
      \frac{1}{4 \pi}
      \rho_{\rm DM} \,  v_0 \, 
          \int_0^{b}  {\rm d} z  \, 2 \pi z \, P_{A'\to \gamma} ( v_r) 
      \nonumber\\
      &=
     P_{A'\to \gamma} ( v_0)
   \, \rho_{\rm DM} \, v(r_c)  \,   \,  r_c^2  \,,
   \label{eq:radpower}
\end{align}
where we consider DM density $\rho_{\rm DM} = 0.4 ~\text{GeV}~\text{cm}^{-3}$ completely composed of 
dark photon.
Its average velocity $ v_0 \simeq 220\, {\rm km/s}$ and the  resonant conversion happens at the solar radius $r_c$.
The parameter $z$ is the impact parameter at infinity for the incoming $A'$, while 
$b$ is the largest value of the impact parameter such that $A'$ can reach the conversion shell at $r
= r_c$. Due to the gravitational focusing enhancement, 
$b = r_c v(r_c)/v_0$ will be larger than $r_c$ in general, by a factor of about 2--3 in numeric calculations. 
The velocity of $A'$ at radius $r_c$ is given by 
$v(r_c) = \sqrt{v_0^2 + 2  G_N M_{\odot}/r_c}$, with $G_N$ being the gravitational constant and $M_{\odot}$ the solar mass.
The radial direction velocity at the conversion point is $v_{r}(z) = \sqrt {  2 {\rm G_N M}_{\rm \odot}/r_c  + v_0^2  - v_0^2 z^2 / r_c^2  }$. 
The factor $2$ in Eq.~\eqref{eq:radpower} counts the DM coming in and going
out of the resonant layer. The converted photon from DM coming in will be
reflected, because when the photon frequency is smaller than the plasma mass, the
total reflection will happen. 

The spectral power flux density emitted per unit solid angle is given as 
\begin{align}
S_{\rm sig} & = \frac{1}{d^2} \frac{1}{\cal B} \frac{d \mathcal{P}}{d \Omega}
\end{align}
where $d = 1 {\rm AU}$ is the distance from the Earth to the Sun,
$ {\cal B} $ is the optimized  bandwidth, which is set as the larger one of the signal bandwidth $B_{\rm sig}$
and the telescope spectral resolution $B_{\rm res}$, namely, ${\cal B} = {\rm max} ( B_{\rm sig}, B_{\rm res})$.
The signal bandwidth $B_{\rm sig}$ is due to the dispersion of the dark photons, 
\begin{equation}
B_{\rm sig} \approx \frac{m_{A'} v_0^2} {2 \pi}\sim 130~ {\rm Hz} \times \frac{m_{A'}}{\rm \mu eV} \,\,,
\label{eq:Bsig}
\end{equation}
which is normally smaller than $B_{\rm res}$. And the telescope spectral resolution $B_{\rm res}$ depends on the 
property of the telescope.
\\

\noindent \textit{\textbf{The Photon Propagation}}-- After the conversion, the propagation of the radio waves in the thermal plasma follows the refraction law, $n\sin\theta = {\rm const}$, where $n$ is the refractive index and $\theta$ is the incident angle. In non-relativistic plasma, $n$ can be expressed as 
\bea
n(\omega) = (1-\omega_p^2/\omega^2)^{1/2} ,
\eea
where $n(\omega)$ equals to the group velocity of the radio waves, i.e. the photon speed. In the resonant region, the dark photon DM 
has a velocity of about $v\sim10^{-3}-10^{-2}$. As a result, the refractive index at the resonant region is in the range of 
$n_{\rm res} \sim 10^{-3}-10^{-2}$, which is much smaller than one. From Fig.~\ref{fig:ne}, the electron density $n_e$ decreases quickly 
with the increase of $r$. Consequently, once the photon leaves the resonant region, the refractive index will quickly go back to 1, 
$n_{\rm out} \sim 1$. Thus, according to the refractive law, the incident angle outside the resonant region can be written as
\bea 
\sin\theta_{\rm out} = \frac{n_{\rm res}}{n_{\rm out}}\times \sin\theta_{\rm res} \lesssim 10^{-3}-10^{-2} \ . 
\eea
Therefore, the direction of the converted photon is approximately along the gradient of the electron density $- \nabla n_e$. 
Considering the conversion happened when the dark photon flies into the Sun and the converted photon moving into the denser region, we expect that electromagnetic waves is always total reflected away from the region where $\omega < \omega_p$. Hence the above discussion 
of the final photon direction applies after the total reflection.
If the electron distribution in the Sun's corona is spherical, the converted radio waves will all propagate along with the radial direction of 
the Sun.
In this case, all the converted radio waves observed on the Earth's surface are from the center of the solar plate. 
However, there are turbulences and flares in the Sun's corona, which makes $n_e$ non-spherical and even evolve with time. 
It will affect the gradient direction of $n_e$, thus modify the out-going direction of the photon.
However, such modification should not have preferred directions, unless there are underline substructures.  
Therefore, we ignore those modifications and assume that in average, the out-going converted photons are isotropic.

Once converted, the radio waves can be absorbed or scattered in the plasma, which is characterized by opacity. 
It turns out that the dominant absorption process is the inverse bremsstrahlung process.
In the corona sphere, the temperature is as high as $10^6 ~{\rm K}$, which is much larger than the ionization energy of the hydrogen atom. As a result, 
the Born approximation can be used to calculate the absorption rate. Since we are interested in the radio wave frequency, 
it satisfies $\omega \ll T\ll m_e$. The absorption rate of the inverse bremsstrahlung process can be calculated as
\bea\label{eq:brem}
\Gamma_{\rm inv} &\approx& \frac{8\pi n_e n_N \alpha^3}{3 \omega^3 m_e^2} \left( \frac{2\pi m_e}{T} \right)^{1/2} \log\left( \frac{2T^2}{\omega_p^2} \right) \left(1 - e^{-\omega/T}\right) \ , \nonumber \\
\eea
where the singularity at $\omega = 0$ clearly shows the effect of the infrared enhancement. $n_N $ is the number density of charged ions.
The logarithmic factor is 
from the long-range effect of the Coulomb interaction, which is cut-off by the Debye screening effect. 
The factor $(1- e^{-\omega/T})$ is due to the stimulated radiation. 
The above calculation is in good agreement with Ref.~\cite{Redondo:2008aa}, except for a minor difference
in the argument of the logarithmic factor. 

Besides the inverse bremsstrahlung process, there is also a contribution from the Compton scattering with the rate given as 
\begin{eqnarray}
\label{eq:Compton}
\Gamma_{\rm Com} &= & \frac{8\pi  \alpha^2}{3  m_e^2}  n_e .
\end{eqnarray}
The Compton scattering can shift the photon energy by a few percent due to the velocity of the electrons. This change is normally larger than the optimized bandwidth. As a conservative consideration,   
we add up the two contributions and have the attenuation rate $\Gamma_{\rm att} = \Gamma_{\rm inv} + \Gamma_{\rm Com}$ for the converted photon.
Numerically, the inverse bremsstrahlung dominates. 
The survival probability $P_s$ for the converted photons to escape the Sun is to add the two rates,
\begin{align}
P_s \equiv  e^{ - \int  { \Gamma_{\rm att} dt}} \simeq \exp\left(- \int_{r_c}^{r_{\max}} \Gamma_{\rm att} dr/v_r \right) \,,
\label{eq:Ps}
\end{align} 
which represents the chance of the photons being not scattered or absorbed during the propagation. 
We terminate the integration at $r_{\max}=10^6~{\rm km} + r_{\rm ps}$ due to 
the available electron density data \cite{2008GeofI..47..197D}, where $r_{\rm ps} =695,510 \, {\rm km}$ is the photosphere radius. 
Further extending the range will not change the result significantly, because the 
electron density is too low such that the interaction rate is negligible.

Dark photon DM with mass $> 4\times10^{-6}$ eV can also convert resonantly to photons in the Sun's chromosphere. However, the temperature of the chromosphere is only about $10^3$ K, which is about three orders of magnitude smaller than the temperature of the 
corona. This makes the inverse bremsstrahlung absorption  much stronger in the chromosphere than in the corona. Furthermore, the 
electron number density, as shown in Fig.~\ref{fig:ne}, is also orders of magnitude larger, and so does the density of charged ions. 
Therefore, the radio waves produced in the chromosphere cannot propagate out. 

In summary, the dark photon DM's resonant conversion happening in the Sun's corona can propagate to the Earth's surface. 
In terms of distance, the region $2300$ km above the photosphere 
(higher than the solar transition region) is our signal region. This corresponds to the unshaded region in Fig.~\ref{fig:ne}.
The relevant observed photon frequency is $\lesssim 1000$ MHz and dark photon mass is $m_{A' } \lesssim 4 \times 10^{-6} \, {\rm  eV}$.
In the above discussions, we only use the well-accepted electron density and temperature profiles as shown in Fig.~\ref{fig:ne}.
They are good approximations and have acceptable uncertainties for the signal calculation.
More discussions on the solar models and the corresponding uncertainties are given in the \textit{Supplemental Material}~\cite{1981ApJS...45..635V, Aschwanden_2001, 1976RSPTA.281..339G, peter1990solar, aschwanden2006physics,1990ApJ...355..700F, 1996ApJS..106..143B}.
\\

\noindent \textit{\textbf{The sensitivity of Radio Telescopes}}--
The minimum detectable flux density of a radio telescope is~\cite{SKA1-Baseline}
\begin{align}
S_{\min} = \frac{\rm SEFD }{\eta_s \sqrt{n_{\rm pol}  \, {\cal B} ~t_{\rm obs}} }\ ,
\label{eq:Smin1}
\end{align}
where $n_{\rm pol} =2$ is the number of polarization, 
$t_{\rm obs}$ is the observation time,
and $\eta_s$ is the system efficiency. In our analysis, we take $\eta_s = 0.9$ for SKA~\cite{SKA1-Baseline}, and 
$\eta_s = 1 $ for LOFAR~\cite{Nijboer:2013dxa}. 
The values of the telescope spectral resolution $B_{\rm res}$ for LOFAR and SKA are listed in Table~\ref{tab:AeffTsys}, which are 
much larger than the signal bandwidth $B_{\rm sig}$ given in Eq.~\eqref{eq:Bsig}.
Therefore, in our calculation, we always have ${\cal B} \simeq  B_{\rm res}$. 
In Eq.~\eqref{eq:Smin1}, SEFD is the system equivalent flux density, defined as
\begin{align}
{\rm SEFD } = 2 k_B \frac{T_{\rm sys} + T_{\odot}^{\rm nos} }{ A_{\rm eff}},
\label{eq:SEFD}
\end{align}
where $k_B$ is the Boltzmann constant, $T_{\rm sys}$ is the antenna system temperature,
$A_{\rm eff}$ is the antenna effective area of the array, and $T_{\odot}^{\rm nos} $ is the antenna noise
temperature increase when pointing to the Sun.

We propose to use the radio telescope arrays SKA and LOFAR to search for the radio waves converted from dark photon DM at the Sun's corona. 
We consider SKA phase 1 (SKA1) as the benchmark of a future telescope to study the reach of dark photon DM. It has a low-frequency aperture 
array (SKA1-Low) and a middle frequency aperture array (SKA1-Mid)~\cite{SKA1-Baseline}.
SKA1-Low
covers the $(50,\,350)$~MHz frequency band. 
SKA1-Mid covers six frequency bands with frequency ranges $(350,\,1050)$~MHz, $(950,\,1760)$~MHz, $(1650,\,3050)$~MHz, $(2800,\,5180)$~MHz,
$(4600,\,8500)$~MHz, and $(8300,\,15300)$~MHz.
In this analysis, to partially cover the frequency range of the converted radio wave, we use the SKA-Low and the first two frequency bands of SKA-Mid, denoted as Mid B1 and Mid B2, respectively.
LOFAR, as an existing radio telescope, can be used for dark photon hunting as well.
Indeed, one of the key science projects for LOFAR is to study solar physics. In its radio spectrometer mode, the intensity of the solar
radio radiation over time is recorded. 
LOFAR covers the frequency ranges of $( 10, \,80 ) \, {\rm MHz}$ and $( 120, \,240 ) \, {\rm MHz}$. 

To calculate the minimum detectable flux $S_{\rm min}$ given in  Eq.~\eqref{eq:Smin1}, we need to determine the corresponding detector parameters, such as 
the telescope spectral resolution $B_{\rm res}$, the system temperature $T_{\rm sys}$, the solar noise temperature $T_{\odot}^{\rm nos} $
and the effective area $A_{\rm eff}$. Table~\ref{tab:AeffTsys} lists the average values of these parameters for each telescope, and the details to achieve these parameters are given as follows:
\begin{table}
	\centering
	\begin{ruledtabular}
		\begin{tabular}{l|l|c|c|c}
		Name	&$f$ [MHz] 
            & $B_{\rm res}$ [kHz] 
            & $\langle T_{\rm sys} \rangle$ [K] 
            & $\langle A_{\rm eff} \rangle$ [${\rm m^2}$] 
               \\
      \hline
		SKA1-Low & (50, 350) & 1 & 680 &  $2.2\times 10^5$ \\
		SKA1-Mid B1 & (350, 1050)  & 3.9 & 28 & $2.7\times 10^4 $ 
         \\
		SKA1-Mid B2 & (950, 1760) &  3.9& 20 & $3.5\times 10^4$  \\
		LOFAR & (10, 80) & 195& 28,110 & 1,830 \\
		LOFAR & (120, 240)& 195& 1,770 & 1,530 \\
		\end{tabular}
	\end{ruledtabular}
	\caption{The frequency range, telescope spectral resolution  $B_{\rm res}$, 
         averaged system temperature $ T_{\rm sys}$ and
         averaged effective area $A_{\rm eff}$
      in the different frequency bands for SKA1 and LOFAR.}
	\label{tab:AeffTsys}
\end{table}

\begin{itemize}[leftmargin=*]
\item
{\it spectral resolution} $B_{\rm res}$:
due to $2.5 \times 10^5$ fine frequency channels in SKA1-Low,
its channel bandwidth can reach $B_{\rm res} = 1$ kHz, 
while the bandwidth for SKA1-Mid B1 and SKA1-Mid B2 are set to $B_{\rm res} = 3.9$ kHz
\cite{SKA1-Baseline}. 
For LOFAR, The spectral resolution $B_{\rm res}$ is taken as $195$ kHz \cite{vanHaarlem:2013dsa, 2019NatAs...3..452M}.

\item {\it system temperature} $T_{\rm sys}$ and {\it effective area} $A_{\rm eff}$:  
the system temperature for SKA1-Low can be approximated as
$T_{\rm sys}^{\rm Low} \approx T_{\rm rec} + T_{\rm sky}$, where the sky noise $T_{\rm sky} \approx 1.23\times 10^8 {\rm K} ~ ({\rm MHz}/f)^{2.55}$ and the receiver noise $T_{\rm rec} = 40 ~{\rm K} + 0.1 T_{\rm sky}$ \cite{SKA1-Baseline}. 
For SKA1-Mid, the average system temperatures for bands 1--5 are $28 {\rm ~K}$, $20 {\rm ~K}$,
$20 {\rm ~K}$, $22 {\rm ~K}$, and $25 {\rm ~K}$, respectively \cite{SKA1-Baseline}.
The effective area $A_{\rm eff}$ is derived using the  system sensitivity $A_{\rm eff}/T_{\rm sys}$ in \cite{vanHaarlem:2013dsa}.
The parameters of LOFAR like
$A_{\rm eff}$ can be directly found in \cite{vanHaarlem:2013dsa}, while $T_{\rm sys}$ can be inferred from SEFD.
Note that in the numeric calculation, the parameters $A_{\rm eff}$ and $T_{\rm sys}$ depend on the frequency.

\item {\it solar noise temperature} $T_{\odot}^{\rm nos} $:
$T_{\odot}^{\rm nos} $ can be calculated under the blackbody assumption for a quiet Sun~\cite{2007IPNPR.168E...1H, 2008IPNPR.175E...1H}. 
The ratio of $T_{\odot}^{\rm nos}$ and the brightness temperature of the quiet Sun $T_b$, 
$T_{\odot}^{\rm nos}/T_b $, has been given for different half-power beamwidth (HPBW or $\text{-3dB}$ beam width) and beam pointing 
offset. 
It is easy to understand that the ratio should always be smaller than 1, because the noise temperature cannot be higher than the 
source itself. 
The result shows that for the antenna with HPBW smaller than the angular diameter of the Sun disk, this ratio is close to 1 when
beam is on the solar disk. The HPBW for SKA1-Low current design is about 4 arcminutes at the baseline frequency 
110~MHz \cite{2015PhDT.......574S}, while the angular diameter of the Sun is as large as 31.8 arcminutes. Therefore, SKA1-Low can 
be considered 
as a high-gain antenna with a very narrow beam. The SKA1-Mid has even smaller HPBW than SKA1-Low, thus
throughout the calculation, we take 
$T_{\odot}^{\rm nos} = T_b$. The spectral brightness temperature $T_b(f)$ is calculated using the  quiet Sun flux density from 
\cite{1986Kraus.book, 2008IPNPR.175E...1H}.
Regarding the LOFAR beamwidth, the HPBW of LOFAR ranges from $(1.3,~19)$ degrees \cite{vanHaarlem:2013dsa}.
Therefore, it is 
much larger than the angular diameter of the Sun. Following the procedure of \cite{2008IPNPR.175E...1H}, 
we use the antenna diameters of LOFAR to
calculate the ratio $T_{\odot}^{\rm nos}/T_{\rm b}$ 
for the Sun as a function of frequency. This ratio is far smaller than one because much of the photon flux goes outside the HPBW.
It is important to remark that this ratio should also work for the signals because both background and signal emissions are 
originated from the Sun.
We find that for the frequency smaller than $55$ MHz, the system temperature $T_{\rm sys}$ dominates over the solar contribution 
$T_{\odot}^{\rm nos}$.
\end{itemize}

\begin{figure}
	\centering
	\includegraphics[width=0.99 \columnwidth]{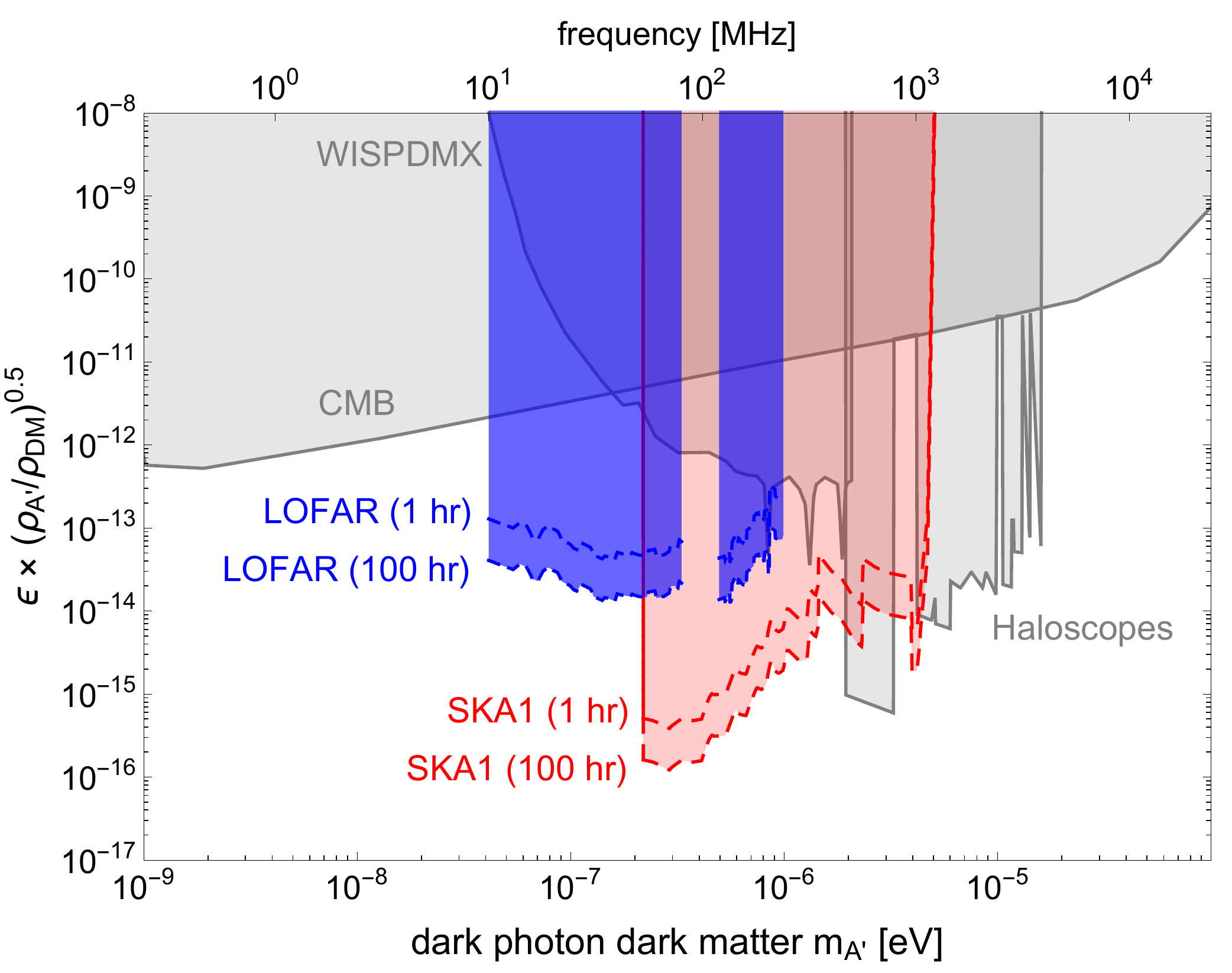} 	
	\caption{
		The sensitivity reach of dark photon dark matter for LOFAR (blue) and SKA1 (red) telescopes with 1 or 100 hours solar observations.
		The constraints are obtained from the existing haloscope axion searches \cite{
			DePanfilis:1987dk, Wuensch:1989sa, Hagmann:1990tj, Asztalos:2001tf, Asztalos:2009yp, Arias:2012az}, 
		recent WISPDMX dark photon searches \cite{Nguyen:2019xuh} and
		the CMB distortion \cite{Arias:2012az, McDermott:2019lch}.
		For both signal and existing constraints, $\rho_{\rm DM} = \rho_{A'}$ is assumed.
	}	\label{fig:final-plot}
\end{figure}

\noindent \textit{\textbf{Results and Discussions}}--
Requiring $S_{\rm sig} \times P_{s} = S_{\min}$, one can obtain the sensitivities on the kinetic mixing $\epsilon$ from radio telescopes.
The sensitivity reaches of dark photon DM for SKA and LOFAR are given in Fig.~\ref{fig:final-plot},
where both the signal and constraints are plotted under the assumption $\rho_{\rm DM} = \rho_{A'}$.
The blue regions show the physics potential of LOFAR with 1-hour and 100-hours observation time, 
which is $1$--$2$ orders of magnitude better than the existing limits from haloscope limits \cite{
DePanfilis:1987dk, Wuensch:1989sa, Hagmann:1990tj, Asztalos:2001tf, Asztalos:2009yp, Arias:2012az}, 
recent WISPDMX constraint \cite{Nguyen:2019xuh} and the CMB distortion \cite{Arias:2012az, McDermott:2019lch}. 
SKA1 has smaller $T_{\rm sys}$, larger $A_{\rm eff}$, and better spectral resolution $B_{\rm res}$. 
Its sensitivities with 1-hour and 100-hours observation time are shown in the red shaded region.
With the same operation time, it can improve the reach of $\epsilon$ by another one or two orders of magnitude compared with LOFAR.

In conclusion, we propose to search for the radiofrequency dark photon DM from $10 -1000\,{\rm MHz}$,
with radio telescopes. In this frequency regime, we show that the dark photon DM can convert resonantly into monochromatic 
radio waves in the solar corona.
In this mass window, the existing LOFAR telescope can achieve a sensitivity of $\sim 10^{-13}-10^{-14}$ on the kinetic mixing $\epsilon$, and the planned SKA1 can achieve a sensitivity of $\sim 10^{-14} - 10^{-16}$.
Despite SKA and LOFAR, other radio telescopes that may be used in the dark photon DM search are MWA~\cite{2005SPIE.5901..124S}, Arecibo~\cite{Giovanelli:2005ee}, JVLA~\cite{VLA}  and FAST~\cite{Nan:2011um}.
In future, the SKA phase 2~\cite{SKA1-tech} can further improve the SEFD sensitivity to sub $\mu$Jy and 
explore more parameter space of the dark photon DM.

\begin{acknowledgments}
The authors would like to thank Goerge Heald, Judith Irwin, Ben Safdi, Lijing Shao, David Tanner, Aaron Vincent and Yiming Zhong for helpful discussions.
The authors would like to express a special thanks to the Mainz Institute for Theoretical Physics (MITP) of the Cluster of Excellence PRISMA+ (Project ID 39083149) workshop for their hospitality and support. 
HA and WX thank the Erwin Schr\"odinger International Institute for hospitality during the completion of this work.
The work of HA is supported by NSFC under Grant No. 11975134, the National Key Research and Development Program of China under Grant No.2017YFA0402204 and the Tsinghua University Initiative Scientific Research Program. FPH is supported by the McDonnell Center for the Space Sciences. The work of JL is supported by NSFC under Grant No. 12075005
and by Peking University under startup Grant No. 7101502458. And the work of WX is supported by the DOE grant DE-SC0010296. 
\end{acknowledgments}

\bibliographystyle{utphys}
\bibliography{ref}

\providecommand{\href}[2]{#2}\begingroup\raggedright\begin{thebibliography}{10}

\bibitem{Essig:2013lka}
R.~Essig {\em et~al.}, ``{Working Group Report: New Light Weakly Coupled
  Particles},'' in {\em {Proceedings, 2013 Community Summer Study on the Future
  of U.S. Particle Physics: Snowmass on the Mississippi (CSS2013): Minneapolis,
  MN, USA, July 29-August 6, 2013}}.
\newblock 2013.
\newblock \href{http://arxiv.org/abs/1311.0029}{{\ttfamily arXiv:1311.0029
  [hep-ph]}}.
\newblock
\url{http://www.slac.stanford.edu/econf/C1307292/docs/IntensityFrontier/NewLight-17.pdf}.
\newblock

\bibitem{Battaglieri:2017aum}
M.~Battaglieri {\em et~al.}, ``{US Cosmic Visions: New Ideas in Dark Matter
  2017: Community Report},'' in {\em {U.S. Cosmic Visions: New Ideas in Dark
  Matter College Park, MD, USA, March 23-25, 2017}}.
\newblock 2017.
\newblock \href{http://arxiv.org/abs/1707.04591}{{\ttfamily arXiv:1707.04591
  [hep-ph]}}.
\newblock
\url{http://lss.fnal.gov/archive/2017/conf/fermilab-conf-17-282-ae-ppd-t.pdf}.
\newblock

\bibitem{Holdom:1985ag}
B.~Holdom, ``{Two U(1)'s and Epsilon Charge Shifts},''
\href{http://dx.doi.org/10.1016/0370-2693(86)91377-8}{{\em Phys. Lett.}
  {\bfseries 166B} (1986) 196--198}.

\bibitem{Redondo:2008ec}
J.~Redondo and M.~Postma, ``{Massive hidden photons as lukewarm dark matter},''
  \href{http://dx.doi.org/10.1088/1475-7516/2009/02/005}{{\em JCAP} {\bfseries
  02} (2009) 005}, \href{http://arxiv.org/abs/0811.0326}{{\ttfamily
  arXiv:0811.0326 [hep-ph]}}.

\bibitem{Nelson:2011sf}
A.~E. Nelson and J.~Scholtz, ``{Dark Light, Dark Matter and the Misalignment
  Mechanism},'' \href{http://dx.doi.org/10.1103/PhysRevD.84.103501}{{\em Phys.
  Rev.} {\bfseries D84} (2011) 103501},
\href{http://arxiv.org/abs/1105.2812}{{\ttfamily arXiv:1105.2812 [hep-ph]}}.

\bibitem{Arias:2012az}
P.~Arias, D.~Cadamuro, M.~Goodsell, J.~Jaeckel, J.~Redondo, and A.~Ringwald,
  ``{WISPy Cold Dark Matter},''
  \href{http://dx.doi.org/10.1088/1475-7516/2012/06/013}{{\em JCAP} {\bfseries
  1206} (2012) 013},
\href{http://arxiv.org/abs/1201.5902}{{\ttfamily arXiv:1201.5902 [hep-ph]}}.

\bibitem{Graham:2015rva}
P.~W. Graham, J.~Mardon, and S.~Rajendran, ``{Vector Dark Matter from
  Inflationary Fluctuations},''
  \href{http://dx.doi.org/10.1103/PhysRevD.93.103520}{{\em Phys. Rev.}
  {\bfseries D93} no.~10, (2016) 103520},
\href{http://arxiv.org/abs/1504.02102}{{\ttfamily arXiv:1504.02102 [hep-ph]}}.

\bibitem{Dienes:1996zr}
K.~R. Dienes, C.~F. Kolda, and J.~March-Russell, ``{Kinetic mixing and the
  supersymmetric gauge hierarchy},''
  \href{http://dx.doi.org/10.1016/S0550-3213(97)00173-9}{{\em Nucl. Phys. B}
  {\bfseries 492} (1997) 104--118},
  \href{http://arxiv.org/abs/hep-ph/9610479}{{\ttfamily arXiv:hep-ph/9610479}}.

\bibitem{Abel:2003ue}
S.~A. Abel and B.~W. Schofield, ``{Brane anti-brane kinetic mixing,
  millicharged particles and SUSY breaking},''
  \href{http://dx.doi.org/10.1016/j.nuclphysb.2004.02.037}{{\em Nucl. Phys. B}
  {\bfseries 685} (2004) 150--170},
  \href{http://arxiv.org/abs/hep-th/0311051}{{\ttfamily arXiv:hep-th/0311051}}.

\bibitem{Abel:2006qt}
S.~A. Abel, J.~Jaeckel, V.~V. Khoze, and A.~Ringwald, ``{Illuminating the
  Hidden Sector of String Theory by Shining Light through a Magnetic Field},''
  \href{http://dx.doi.org/10.1016/j.physletb.2008.03.076}{{\em Phys. Lett. B}
  {\bfseries 666} (2008) 66--70},
  \href{http://arxiv.org/abs/hep-ph/0608248}{{\ttfamily arXiv:hep-ph/0608248}}.

\bibitem{Abel:2008ai}
S.~A. Abel, M.~D. Goodsell, J.~Jaeckel, V.~V. Khoze, and A.~Ringwald,
  ``{Kinetic Mixing of the Photon with Hidden U(1)s in String Phenomenology},''
  \href{http://dx.doi.org/10.1088/1126-6708/2008/07/124}{{\em JHEP} {\bfseries
  07} (2008) 124}, \href{http://arxiv.org/abs/0803.1449}{{\ttfamily
  arXiv:0803.1449 [hep-ph]}}.

\bibitem{Goodsell:2009xc}
M.~Goodsell, J.~Jaeckel, J.~Redondo, and A.~Ringwald, ``{Naturally Light Hidden
  Photons in LARGE Volume String Compactifications},''
  \href{http://dx.doi.org/10.1088/1126-6708/2009/11/027}{{\em JHEP} {\bfseries
  11} (2009) 027}, \href{http://arxiv.org/abs/0909.0515}{{\ttfamily
  arXiv:0909.0515 [hep-ph]}}.

\bibitem{Lobanov:2012pt}
A.~P. Lobanov, H.~S. Zechlin, and D.~Horns, ``{Astrophysical searches for a
  hidden-photon signal in the radio regime},''
  \href{http://dx.doi.org/10.1103/PhysRevD.87.065004}{{\em Phys. Rev.}
  {\bfseries D87} no.~6, (2013) 065004},
\href{http://arxiv.org/abs/1211.6268}{{\ttfamily arXiv:1211.6268
  [astro-ph.CO]}}.

\bibitem{Mirizzi:2009iz}
A.~Mirizzi, J.~Redondo, and G.~Sigl, ``{Microwave Background Constraints on
  Mixing of Photons with Hidden Photons},''
  \href{http://dx.doi.org/10.1088/1475-7516/2009/03/026}{{\em JCAP} {\bfseries
  0903} (2009) 026},
\href{http://arxiv.org/abs/0901.0014}{{\ttfamily arXiv:0901.0014 [hep-ph]}}.

\bibitem{Kunze:2015noa}
K.~E. Kunze and M.~A. V\'azquez-Mozo, ``{Constraints on hidden photons from
  current and future observations of CMB spectral distortions},''
  \href{http://dx.doi.org/10.1088/1475-7516/2015/12/028}{{\em JCAP} {\bfseries
  12} (2015) 028}, \href{http://arxiv.org/abs/1507.02614}{{\ttfamily
  arXiv:1507.02614 [astro-ph.CO]}}.

\bibitem{Dubovsky:2015cca}
S.~Dubovsky and G.~Hernández-Chifflet, ``{Heating up the Galaxy with Hidden
  Photons},'' \href{http://dx.doi.org/10.1088/1475-7516/2015/12/054}{{\em JCAP}
  {\bfseries 1512} no.~12, (2015) 054},
\href{http://arxiv.org/abs/1509.00039}{{\ttfamily arXiv:1509.00039 [hep-ph]}}.

\bibitem{Kovetz:2018zes}
E.~D. Kovetz, I.~Cholis, and D.~E. Kaplan, ``{Bounds on ultralight
  hidden-photon dark matter from observation of the 21 cm signal at cosmic
  dawn},'' \href{http://dx.doi.org/10.1103/PhysRevD.99.123511}{{\em Phys. Rev.
  D} {\bfseries 99} no.~12, (2019) 123511},
  \href{http://arxiv.org/abs/1809.01139}{{\ttfamily arXiv:1809.01139
  [astro-ph.CO]}}.

\bibitem{Pospelov:2018kdh}
M.~Pospelov, J.~Pradler, J.~T. Ruderman, and A.~Urbano, ``{Room for New Physics
  in the Rayleigh-Jeans Tail of the Cosmic Microwave Background},''
  \href{http://dx.doi.org/10.1103/PhysRevLett.121.031103}{{\em Phys. Rev.
  Lett.} {\bfseries 121} no.~3, (2018) 031103},
\href{http://arxiv.org/abs/1803.07048}{{\ttfamily arXiv:1803.07048 [hep-ph]}}.

\bibitem{McDermott:2019lch}
S.~D. McDermott and S.~J. Witte, ``{Cosmological evolution of light dark photon
  dark matter},'' \href{http://dx.doi.org/10.1103/PhysRevD.101.063030}{{\em
  Phys. Rev.} {\bfseries D101} no.~6, (2020) 063030},
\href{http://arxiv.org/abs/1911.05086}{{\ttfamily arXiv:1911.05086 [hep-ph]}}.

\bibitem{Caputo:2020bdy}
A.~Caputo, H.~Liu, S.~Mishra-Sharma, and J.~T. Ruderman, ``{Dark Photon
  Oscillations in Our Inhomogeneous Universe},''
  \href{http://dx.doi.org/10.1103/PhysRevLett.125.221303}{{\em Phys. Rev.
  Lett.} {\bfseries 125} no.~22, (2020) 221303},
  \href{http://arxiv.org/abs/2002.05165}{{\ttfamily arXiv:2002.05165
  [astro-ph.CO]}}.

\bibitem{Garcia:2020qrp}
A.~A. Garcia, K.~Bondarenko, S.~Ploeckinger, J.~Pradler, and A.~Sokolenko,
  ``{Effective photon mass and (dark) photon conversion in the inhomogeneous
  Universe},'' \href{http://dx.doi.org/10.1088/1475-7516/2020/10/011}{{\em
  JCAP} {\bfseries 10} (2020) 011},
  \href{http://arxiv.org/abs/2003.10465}{{\ttfamily arXiv:2003.10465
  [astro-ph.CO]}}.

\bibitem{Baryakhtar:2018doz}
M.~Baryakhtar, J.~Huang, and R.~Lasenby, ``{Axion and hidden photon dark matter
  detection with multilayer optical haloscopes},''
  \href{http://dx.doi.org/10.1103/PhysRevD.98.035006}{{\em Phys. Rev. D}
  {\bfseries 98} no.~3, (2018) 035006},
  \href{http://arxiv.org/abs/1803.11455}{{\ttfamily arXiv:1803.11455
  [hep-ph]}}.

\bibitem{Pospelov:2008jk}
M.~Pospelov, A.~Ritz, and M.~B. Voloshin, ``{Bosonic super-WIMPs as keV-scale
  dark matter},'' \href{http://dx.doi.org/10.1103/PhysRevD.78.115012}{{\em
  Phys. Rev. D} {\bfseries 78} (2008) 115012},
  \href{http://arxiv.org/abs/0807.3279}{{\ttfamily arXiv:0807.3279 [hep-ph]}}.

\bibitem{An:2014twa}
H.~An, M.~Pospelov, J.~Pradler, and A.~Ritz, ``{Direct Detection Constraints on
  Dark Photon Dark Matter},''
  \href{http://dx.doi.org/10.1016/j.physletb.2015.06.018}{{\em Phys. Lett. B}
  {\bfseries 747} (2015) 331--338},
  \href{http://arxiv.org/abs/1412.8378}{{\ttfamily arXiv:1412.8378 [hep-ph]}}.

\bibitem{Bloch:2016sjj}
I.~M. Bloch, R.~Essig, K.~Tobioka, T.~Volansky, and T.-T. Yu, ``{Searching for
  Dark Absorption with Direct Detection Experiments},''
  \href{http://dx.doi.org/10.1007/JHEP06(2017)087}{{\em JHEP} {\bfseries 06}
  (2017) 087}, \href{http://arxiv.org/abs/1608.02123}{{\ttfamily
  arXiv:1608.02123 [hep-ph]}}.

\bibitem{Aprile:2019xxb}
{\bfseries XENON} Collaboration, E.~Aprile {\em et~al.}, ``{Light Dark Matter
  Search with Ionization Signals in XENON1T},''
  \href{http://dx.doi.org/10.1103/PhysRevLett.123.251801}{{\em Phys. Rev.
  Lett.} {\bfseries 123} no.~25, (2019) 251801},
  \href{http://arxiv.org/abs/1907.11485}{{\ttfamily arXiv:1907.11485
  [hep-ex]}}.

\bibitem{An:2013yfc}
H.~An, M.~Pospelov, and J.~Pradler, ``{New stellar constraints on dark
  photons},'' \href{http://dx.doi.org/10.1016/j.physletb.2013.07.008}{{\em
  Phys. Lett. B} {\bfseries 725} (2013) 190--195},
  \href{http://arxiv.org/abs/1302.3884}{{\ttfamily arXiv:1302.3884 [hep-ph]}}.

\bibitem{Redondo:2013lna}
J.~Redondo and G.~Raffelt, ``{Solar constraints on hidden photons
  re-visited},'' \href{http://dx.doi.org/10.1088/1475-7516/2013/08/034}{{\em
  JCAP} {\bfseries 1308} (2013) 034},
\href{http://arxiv.org/abs/1305.2920}{{\ttfamily arXiv:1305.2920 [hep-ph]}}.

\bibitem{Vinyoles:2015aba}
N.~Vinyoles, A.~Serenelli, F.~L. Villante, S.~Basu, J.~Redondo, and J.~Isern,
  ``{New axion and hidden photon constraints from a solar data global fit},''
  \href{http://dx.doi.org/10.1088/1475-7516/2015/10/015}{{\em JCAP} {\bfseries
  1510} (2015) 015},
\href{http://arxiv.org/abs/1501.01639}{{\ttfamily arXiv:1501.01639
  [astro-ph.SR]}}.

\bibitem{An:2020bxd}
H.~An, M.~Pospelov, J.~Pradler, and A.~Ritz, ``{New limits on dark photons from
  solar emission and keV scale dark matter},''
  \href{http://arxiv.org/abs/2006.13929}{{\ttfamily arXiv:2006.13929
  [hep-ph]}}.

\bibitem{An:2013yua}
H.~An, M.~Pospelov, and J.~Pradler, ``{Dark Matter Detectors as Dark Photon
  Helioscopes},'' \href{http://dx.doi.org/10.1103/PhysRevLett.111.041302}{{\em
  Phys. Rev. Lett.} {\bfseries 111} (2013) 041302},
\href{http://arxiv.org/abs/1304.3461}{{\ttfamily arXiv:1304.3461 [hep-ph]}}.

\bibitem{She:2019skm}
{\bfseries CDEX} Collaboration, Z.~She {\em et~al.}, ``{Direct Detection
  Constraints on Dark Photons with the CDEX-10 Experiment at the China Jinping
  Underground Laboratory},''
  \href{http://dx.doi.org/10.1103/PhysRevLett.124.111301}{{\em Phys. Rev.
  Lett.} {\bfseries 124} no.~11, (2020) 111301},
  \href{http://arxiv.org/abs/1910.13234}{{\ttfamily arXiv:1910.13234
  [hep-ex]}}.

\bibitem{Ema:2019yrd}
Y.~Ema, K.~Nakayama, and Y.~Tang, ``{Production of Purely Gravitational Dark
  Matter: The Case of Fermion and Vector Boson},''
  \href{http://dx.doi.org/10.1007/JHEP07(2019)060}{{\em JHEP} {\bfseries 07}
  (2019) 060}, \href{http://arxiv.org/abs/1903.10973}{{\ttfamily
  arXiv:1903.10973 [hep-ph]}}.

\bibitem{Co:2018lka}
R.~T. Co, A.~Pierce, Z.~Zhang, and Y.~Zhao, ``{Dark Photon Dark Matter Produced
  by Axion Oscillations},''
\href{http://arxiv.org/abs/1810.07196}{{\ttfamily arXiv:1810.07196 [hep-ph]}}.

\bibitem{Dror:2018pdh}
J.~A. Dror, K.~Harigaya, and V.~Narayan, ``{Parametric Resonance Production of
  Ultralight Vector Dark Matter},''
\href{http://arxiv.org/abs/1810.07195}{{\ttfamily arXiv:1810.07195 [hep-ph]}}.

\bibitem{Bastero-Gil:2018uel}
M.~Bastero-Gil, J.~Santiago, L.~Ubaldi, and R.~Vega-Morales, ``{Vector dark
  matter production at the end of inflation},''
\href{http://arxiv.org/abs/1810.07208}{{\ttfamily arXiv:1810.07208 [hep-ph]}}.

\bibitem{Agrawal:2018vin}
P.~Agrawal, N.~Kitajima, M.~Reece, T.~Sekiguchi, and F.~Takahashi, ``{Relic
  Abundance of Dark Photon Dark Matter},''
\href{http://arxiv.org/abs/1810.07188}{{\ttfamily arXiv:1810.07188 [hep-ph]}}.

\bibitem{Long:2019lwl}
A.~J. Long and L.-T. Wang, ``{Dark Photon Dark Matter from a Network of Cosmic
  Strings},''
\href{http://arxiv.org/abs/1901.03312}{{\ttfamily arXiv:1901.03312 [hep-ph]}}.

\bibitem{AlonsoAlvarez:2019cgw}
G.~Alonso-Álvarez, T.~Hugle, and J.~Jaeckel, ``{Misalignment \& Co.:
  (Pseudo-)scalar and vector dark matter with curvature couplings},''
\href{http://arxiv.org/abs/1905.09836}{{\ttfamily arXiv:1905.09836 [hep-ph]}}.

\bibitem{Nakayama:2019rhg}
K.~Nakayama, ``{Vector Coherent Oscillation Dark Matter},''
  \href{http://dx.doi.org/10.1088/1475-7516/2019/10/019}{{\em JCAP} {\bfseries
  1910} (2019) 019},
\href{http://arxiv.org/abs/1907.06243}{{\ttfamily arXiv:1907.06243 [hep-ph]}}.

\bibitem{Nakai:2020cfw}
Y.~Nakai, R.~Namba, and Z.~Wang, ``{Light Dark Photon Dark Matter from
  Inflation},''
\href{http://arxiv.org/abs/2004.10743}{{\ttfamily arXiv:2004.10743 [hep-ph]}}.

\bibitem{DePanfilis:1987dk}
S.~De~Panfilis, A.~C. Melissinos, B.~E. Moskowitz, J.~T. Rogers, Y.~K.
  Semertzidis, W.~Wuensch, H.~J. Halama, A.~G. Prodell, W.~B. Fowler, and F.~A.
  Nezrick, ``{Limits on the Abundance and Coupling of Cosmic Axions at
  4.5-Microev < m(a) < 5.0-Microev},''
\href{http://dx.doi.org/10.1103/PhysRevLett.59.839}{{\em Phys. Rev. Lett.}
  {\bfseries 59} (1987) 839}.

\bibitem{Wuensch:1989sa}
W.~Wuensch, S.~De~Panfilis-Wuensch, Y.~K. Semertzidis, J.~T. Rogers, A.~C.
  Melissinos, H.~J. Halama, B.~E. Moskowitz, A.~G. Prodell, W.~B. Fowler, and
  F.~A. Nezrick, ``{Results of a Laboratory Search for Cosmic Axions and Other
  Weakly Coupled Light Particles},''
\href{http://dx.doi.org/10.1103/PhysRevD.40.3153}{{\em Phys. Rev.} {\bfseries
  D40} (1989) 3153}.

\bibitem{Hagmann:1990tj}
C.~Hagmann, P.~Sikivie, N.~S. Sullivan, and D.~B. Tanner, ``{Results from a
  search for cosmic axions},''
\href{http://dx.doi.org/10.1103/PhysRevD.42.1297}{{\em Phys. Rev.} {\bfseries
  D42} (1990) 1297--1300}.

\bibitem{Asztalos:2001tf}
{\bfseries ADMX} Collaboration, S.~J. Asztalos {\em et~al.}, ``{Large scale
  microwave cavity search for dark matter axions},''
\href{http://dx.doi.org/10.1103/PhysRevD.64.092003}{{\em Phys. Rev.} {\bfseries
  D64} (2001) 092003}.

\bibitem{Asztalos:2009yp}
{\bfseries ADMX} Collaboration, S.~J. Asztalos {\em et~al.}, ``{A SQUID-based
  microwave cavity search for dark-matter axions},''
  \href{http://dx.doi.org/10.1103/PhysRevLett.104.041301}{{\em Phys. Rev.
  Lett.} {\bfseries 104} (2010) 041301},
\href{http://arxiv.org/abs/0910.5914}{{\ttfamily arXiv:0910.5914
  [astro-ph.CO]}}.

\bibitem{Nguyen:2019xuh}
L.~Hoang~Nguyen, A.~Lobanov, and D.~Horns, ``{First results from the WISPDMX
  radio frequency cavity searches for hidden photon dark matter},''
  \href{http://dx.doi.org/10.1088/1475-7516/2019/10/014}{{\em JCAP} {\bfseries
  1910} no.~10, (2019) 014},
\href{http://arxiv.org/abs/1907.12449}{{\ttfamily arXiv:1907.12449 [hep-ex]}}.

\bibitem{Horns:2012jf}
D.~Horns, J.~Jaeckel, A.~Lindner, A.~Lobanov, J.~Redondo, and A.~Ringwald,
  ``{Searching for WISPy Cold Dark Matter with a Dish Antenna},''
  \href{http://dx.doi.org/10.1088/1475-7516/2013/04/016}{{\em JCAP} {\bfseries
  1304} (2013) 016},
\href{http://arxiv.org/abs/1212.2970}{{\ttfamily arXiv:1212.2970 [hep-ph]}}.

\bibitem{Knirck:2018ojz}
S.~Knirck, T.~Yamazaki, Y.~Okesaku, S.~Asai, T.~Idehara, and T.~Inada, ``{First
  results from a hidden photon dark matter search in the meV sector using a
  plane-parabolic mirror system},''
  \href{http://dx.doi.org/10.1088/1475-7516/2018/11/031}{{\em JCAP} {\bfseries
  1811} no.~11, (2018) 031},
\href{http://arxiv.org/abs/1806.05120}{{\ttfamily arXiv:1806.05120 [hep-ex]}}.

\bibitem{Gelmini:2020kcu}
G.~B. Gelmini, A.~J. Millar, V.~Takhistov, and E.~Vitagliano, ``{Probing dark
  photons with plasma haloscopes},''
  \href{http://dx.doi.org/10.1103/PhysRevD.102.043003}{{\em Phys. Rev. D}
  {\bfseries 102} no.~4, (2020) 043003},
  \href{http://arxiv.org/abs/2006.06836}{{\ttfamily arXiv:2006.06836
  [hep-ph]}}.

\bibitem{vanHaarlem:2013dsa}
M.~P. van Haarlem {\em et~al.}, ``{LOFAR: The LOw-Frequency ARray},''
  \href{http://dx.doi.org/10.1051/0004-6361/201220873}{{\em Astron. Astrophys.}
  {\bfseries 556} (2013) A2},
\href{http://arxiv.org/abs/1305.3550}{{\ttfamily arXiv:1305.3550
  [astro-ph.IM]}}.

\bibitem{SKA1-tech}
S.~collaboration, ``Ska1 info sheets: The telescopes.''
  \url{https://www.skatelescope.org/technical/info-sheets/}, 08, 2018.

\bibitem{Raffelt:1987im}
G.~Raffelt and L.~Stodolsky, ``{Mixing of the Photon with Low Mass
  Particles},''
\href{http://dx.doi.org/10.1103/PhysRevD.37.1237}{{\em Phys. Rev.} {\bfseries
  D37} (1988) 1237}.

\bibitem{2008GeofI..47..197D}
V.~{De La Luz}, A.~{Lara}, E.~{Mendoza}, and M.~{Shimojo}, ``{3D Simulations of
  the Quiet Sun Radio Emission at Millimeter and Submillimeter Wavelengths},''
  {\em Geofisica Internacional} {\bfseries 47} (Jul, 2008) 197--203.

\bibitem{Redondo:2008aa}
J.~Redondo, ``{Helioscope Bounds on Hidden Sector Photons},''
  \href{http://dx.doi.org/10.1088/1475-7516/2008/07/008}{{\em JCAP} {\bfseries
  0807} (2008) 008},
\href{http://arxiv.org/abs/0801.1527}{{\ttfamily arXiv:0801.1527 [hep-ph]}}.

\bibitem{1981ApJS...45..635V}
J.~E. {Vernazza}, E.~H. {Avrett}, and R.~{Loeser}, ``{Structure of the solar
  chromosphere. III. Models of the EUV brightness components of the quiet
  sun.},'' \href{http://dx.doi.org/10.1086/190731}{{\em Astrophysical Journal,
  Suppl. Ser.} {\bfseries 45} (Apr., 1981) 635--725}.

\bibitem{Aschwanden_2001}
M.~J. Aschwanden and L.~W. Acton, ``Tempurature tomography of the soft x-ray
  corona: Measurements of electron densities, tempuratures, and differential
  emission measure distributions above the limb,''
  \href{http://dx.doi.org/10.1086/319711}{{\em The Astrophysical Journal}
  {\bfseries 550} no.~1, (Mar, 2001) 475--492}.
  \url{https://doi.org/10.1086/319711}.

\bibitem{1976RSPTA.281..339G}
A.~H. {Gabriel}, ``{A Magnetic Model of the Solar Transition Region},''
  \href{http://dx.doi.org/10.1098/rsta.1976.0031}{{\em Philosophical
  Transactions of the Royal Society of London Series A} {\bfseries 281}
  no.~1304, (May, 1976) 339--352}.

\bibitem{peter1990solar}
P.~Foukal, {\em Solar Astrophysics}.
\newblock A Wiley-Interscience publication. Wiley, 1990.

\bibitem{aschwanden2006physics}
M.~Aschwanden, {\em Physics of the Solar Corona: An Introduction with Problems
  and Solutions}.
\newblock Springer Praxis Books. Springer Berlin Heidelberg, 2006.
\newblock \url{https://books.google.com/books?id=W7FE5\_aowEQC}.

\bibitem{1990ApJ...355..700F}
J.~M. {Fontenla}, E.~H. {Avrett}, and R.~{Loeser}, ``{Energy Balance in the
  Solar Transition Region. I. Hydrostatic Thermal Models with Ambipolar
  Diffusion},'' \href{http://dx.doi.org/10.1086/168803}{{\em Astrophysical
  Journal} {\bfseries 355} (June, 1990) 700}.

\bibitem{1996ApJS..106..143B}
J.~W. {Brosius}, J.~M. {Davila}, R.~J. {Thomas}, and B.~C. {Monsignori-Fossi},
  ``{Measuring Active and Quiet-Sun Coronal Plasma Properties with
  Extreme-Ultraviolet Spectra from SERTS},''
  \href{http://dx.doi.org/10.1086/192332}{{\em Astrophysical Journal}
  {\bfseries 106} (Sept., 1996) 143}.

\bibitem{SKA1-Baseline}
S.~collaboration, ``Ska1 system baseline design.''
  \url{https://www.skatelescope.org/wp-content/uploads/2014/11/SKA-TEL-SKO-0000002-AG-BD-DD-Rev01-SKA1_System_Baseline_Design.pdf},
  03, 2013.
\newblock Document number: SKA-TEL-SKO-DD-001.

\bibitem{Nijboer:2013dxa}
R.~J. Nijboer, M.~Pandey-Pommier, and A.~G. de~Bruyn, ``{LOFAR imaging
  capabilities and system sensitivity},''
\href{http://arxiv.org/abs/1308.4267}{{\ttfamily arXiv:1308.4267
  [astro-ph.IM]}}.

\bibitem{2019NatAs...3..452M}
D.~E. {Morosan}, E.~P. {Carley}, L.~A. {Hayes}, S.~A. {Murray}, P.~{Zucca},
  R.~A. {Fallows}, J.~{McCauley}, E.~K.~J. {Kilpua}, G.~{Mann}, C.~{Vocks}, and
  P.~T. {Gallagher}, ``{Multiple regions of shock-accelerated particles during
  a solar coronal mass ejection},''
  \href{http://dx.doi.org/10.1038/s41550-019-0689-z}{{\em Nature Astronomy}
  {\bfseries 3} (Feb., 2019) 452--461},
  \href{http://arxiv.org/abs/1908.11743}{{\ttfamily arXiv:1908.11743
  [astro-ph.SR]}}.

\bibitem{2007IPNPR.168E...1H}
C.~{Ho}, A.~{Kantak}, S.~{Slobin}, and D.~{Morabito}, ``{Link Analysis of a
  Telecommunications System on Earth, in Geostationary Orbit, and at the Moon:
  Atmospheric Attenuation and Noise Temperature Effects},'' {\em Interplanetary
  Network Progress Report} {\bfseries 42-168} (Feb., 2007) 1--22.

\bibitem{2008IPNPR.175E...1H}
C.~{Ho}, S.~{Slobin}, A.~{Kantak}, and S.~{Asmar}, ``{Solar Brightness
  Temperature and Corresponding Antenna Noise Temperature at Microwave
  Frequencies},'' {\em Interplanetary Network Progress Report} {\bfseries
  42-175} (Nov., 2008) 1--11.

\bibitem{2015PhDT.......574S}
D.~R. {Sinclair}, {\em {A Study of the Square Kilometre Array Low-Frequency
  Aperture Array}}.
\newblock PhD thesis, University of Oxford, Jan., 2015.

\bibitem{1986Kraus.book}
J.~D. {Kraus}, {\em {Radio astronomy, 2nd edition}}.
\newblock Powell, Ohio: Cygnus-Quasar Books, 1986.

\bibitem{2005SPIE.5901..124S}
J.~E. {Salah}, C.~J. {Lonsdale}, D.~{Oberoi}, R.~J. {Cappallo}, and J.~C.
  {Kasper}, \href{http://dx.doi.org/10.1117/12.613448}{``{Space weather
  capabilities of low frequency radio arrays},''} in {\em Solar Physics and
  Space Weather Instrumentation}, S.~{Fineschi} and R.~A. {Viereck}, eds.,
  vol.~5901 of {\em Society of Photo-Optical Instrumentation Engineers (SPIE)
  Conference Series}, pp.~124--134.
\newblock Aug., 2005.

\bibitem{Giovanelli:2005ee}
R.~Giovanelli {\em et~al.}, ``{The Arecibo Legacy Fast ALFA Survey. 1. Science
  goals, survey design and strategy},''
  \href{http://dx.doi.org/10.1086/497431}{{\em Astron. J.} {\bfseries 130}
  (2005) 2598--2612},
\href{http://arxiv.org/abs/astro-ph/0508301}{{\ttfamily arXiv:astro-ph/0508301
  [astro-ph]}}.

\bibitem{VLA}
{Karl G. Jansky Very Large Array}.
  \url{https://science.nrao.edu/facilities/vla}.

\bibitem{Nan:2011um}
R.~Nan, D.~Li, C.~Jin, Q.~Wang, L.~Zhu, W.~Zhu, H.~Zhang, Y.~Yue, and L.~Qian,
  ``{The Five-Hundred-Meter Aperture Spherical Radio Telescope (FAST)
  Project},'' \href{http://dx.doi.org/10.1142/S0218271811019335}{{\em Int. J.
  Mod. Phys.} {\bfseries D20} (2011) 989--1024},
\href{http://arxiv.org/abs/1105.3794}{{\ttfamily arXiv:1105.3794
  [astro-ph.IM]}}.

\end{thebibliography}\endgroup

\newpage

\onecolumngrid

\fontsize{12pt}{14pt}\selectfont
\setlength{\parindent}{15pt}
\setlength{\parskip}{1em}

\begin{center}
	\textbf{\large Radiofrequency Dark Photon Dark Matter across the Sun} \\ 
	\vspace{0.05in}
	{ \it \large Supplemental Material}\\ 
	\vspace{0.05in}
	{Haipeng An$^{1,2}$, Fa Peng Huang$^{3}$, Jia Liu$^{4,5}$ and Wei Xue$^{6}$}
	\vspace{0.05in}
\end{center}
\centerline{{\it  $^{1}$Department of Physics, Tsinghua University, Beijing 100084, China}}
\centerline{{\it  $^{2}$Center for High Energy Physics, Tsinghua University, Beijing 100084, China}}
\centerline{{\it $^{3}$Department of Physics and McDonnell Center for the Space Sciences, }}
\centerline{{\it Washington University, St. Louis, MO 63130, USA}}
\centerline{{\it  $^{4}$School of Physics and State Key Laboratory of Nuclear Physics and Technology,}}
\centerline{{\it Peking University, Beijing 100871, China}}
\centerline{{\it  $^{5}$Center for High Energy Physics, Peking University, Beijing 100871, China}}
\centerline{{\it  $^{6}$Department of Physics, University of Florida, Gainesville, FL 32611, USA}}
\vspace{0.05in}
{In this supplemental material, we show the derivation of the conversion probability of a dark photon particle $P_{A'\to \gamma}$ for the resonant conversion using quantum field method and linearized wave method. 
We also discuss the solar model we used in the study and compare it with the experimental observations. Lastly, we discuss the uncertainties in the calculation.}
\vspace{0.15in}

\noindent \textbf{The first method--} we use the quantum field method to calculate the $1 \to 1$ conversion rate $\Gamma_{A'\to \gamma}$. We further integrate this rate with the time it takes to fly across the solar corona and obtain the conversion
probability $P_{A'\to \gamma}$. 
\begin{align}
	& P_{A' \to \gamma} ( v_r ) = \int dt \Gamma_{A'\to \gamma} , \\
	& = \int \frac { {\rm d} t }{2 \omega} \frac{ {\rm d}^3 p }{ (2 \pi )^3 2 \omega } 
	( 2\pi)^4 \delta^{4} \left( p_{A'}^\mu - p_\gamma^\mu \right) \,  \frac{1}{3} \sum_{\rm pol}|{ \cal M} |^2  .
	\label{eq:Pcalc}
\end{align}
We take average of the initial dark photon state. The factor $1/3$ is the initial spin average for $A'$.
For the final state, we only sum over the transverse modes.  After the structure formation, the $A'$ dark matter has fallen into the
gravitational well of galaxies and clusters. The gravitational forces changes the momentum of $A'$ together with its direction.
Therefore, we assume in the solar system the $A'$ polarization has equal probability for two transverse modes and one
longitudinal mode. The amplitude $\cal M$ is given as
\begin{align}
	{\cal M} = -\epsilon m_{A'}^2 \left( \xi^*_{\gamma}(p)\cdot \xi_{A'}(p) \right).
\end{align}
For $1\to 1$ process, the energy-momentum conservation implies $p_{A'} = p_{\gamma}\equiv p$.
For photon in the final states, we only count two transverse modes, because longitudinal photon cannot propagate to the Earth.
Therefore, we have 
\begin{align}
	\frac{1}{3} \sum_{\rm pol}|{ \cal M} |^2 = \frac{2}{3} \epsilon^2 m_{A'}^4, 
\end{align}
for the amplitude square. For Eq.~\eqref{eq:Pcalc}, after integrating $d^3p$, there is one $\delta$ function left for energy conservation. Together with the integration of $dt$, it has 
\begin{align}
	\int dt \delta(E_{A'} - E_{\gamma}) = 2 \omega^{-1} \left(\frac{\partial \ln \omega_p^2}{\partial t}\right)^{-1},
\end{align}
where we have used $E_{\gamma} = \sqrt{\vec{p}^2 + \omega_p^2}$ and $\omega_p$ is the plasma frequency which is location dependent. In our assumption, the electron density distribution is spherical symmetric, thus $\omega_p$ only depends on radius $r$. We can further apply $\partial t = v_r^{-1} \partial r$, because only radial movement changes $\omega_p$. Putting all the elements together, we arrive at the final result
\begin{align}
	P_{A' \to \gamma} ( v_r ) = \frac{2}{3} \times 
	{ \pi \, \epsilon^2 \, m_{A'} }
	\, v_r^{-1} \,
	\left| \frac{\partial \ln \omega_p^2 ( r )}  
	{   \partial r     }  \right|^{-1}_{r=r_c} .
	\label{eq:Pfinal}
\end{align}
Since this is $1 \to 1$ process, the momentum conservation requires $ \omega_p(r_c) =m_{A'} $, that the process happens at resonant region $r_c$. \\

\noindent \textbf{The second method--} After the quantum field calculation, we use the linearized wave method to calculate the 
conversion probability. After eliminating the kinetic mixing term by redefinition, one can arrive at the coupled wave 
equations,
\begin{align}
   \left[- \frac{\partial^2}{\partial t^2} + \frac{\partial^2}{\partial{r}^2} - \left(  
	\begin{array}{c c}
		\omega_p^2 & -\epsilon m_{A'}^2 \\
		-\epsilon m_{A'}^2 & m_{A'}^2
 	\end{array}
	\right) \right]
	\left(	\begin{array}{c }
		A(r,t) \\
		A'(r,t)
	\end{array}\right)=0.
\label{eq:wave1}
\end{align} 
We consider solutions with fixed frequency, $\omega$. We define $k = (\omega^2 - m_{A'}^2)^{1/2}$. Then the solution of Eq.~(\ref{eq:wave1}) can be written as $A(r,t) = e^{i(\omega t - r k)}\tilde{A}(r)$ and $A'(r,t) = e^{i(\omega t - r k)}\tilde{A}'(r)$. The plasma frequency is slowly varying compared with the $k$. As a result, we have $|\partial_r \tilde A(r)| \ll k |\tilde A(r)|$ and 
$|\partial_r^2 \tilde A(r)| \ll k |\partial_r \tilde A(r)|$, and the same is true for $A'$ field. Then, we can use the WKB approximation to rewrite Eq.~(\ref{eq:wave1}) as a first-order differential equation, 
\begin{align}
& \left[	-i\partial_r + H_0 + H_I \right] \left(	\begin{array}{c }
	\tilde{A}(r) \\
	\tilde{A}'(r)
\end{array}\right)=0, 
\end{align}
where
\begin{align}
& H_0 = \left(  \begin{array}{c c}
	\frac{m_{A'}^2-\omega_p^2}{2k} & 0 \\
	0 & 0
\end{array} \right), ~
H_I = \left(  \begin{array}{c c}
	0 & -\frac{\epsilon m_{A'}^2}{2k} \\
	-\frac{\epsilon m_{A'}^2}{2k} & 0
\end{array} \right).
\end{align}
Since $H_I$ is much smaller than $H_0$, the first-order solution for the conversion probability is
\begin{align}
P_{A'\to \gamma}= 	\left| \int_{0}^{\infty}d r \frac{-\epsilon m_{A'}^2}{2 k} e^{-i \int_0^{r}d \tilde{r} \frac{m_{A'}^2 - \omega_p^2(\tilde{r})}{2k} }\right|^2 .
\label{eq:Pfinal2}
\end{align}
The result can be further simplified using the \textit{saddle point approximation},
\begin{align}
	\int^{\infty}_{-\infty} dr e^{-f(r ) }\approx e^{-f(r_0)} \sqrt{\frac{2 \pi}{f^{''}(r_0)}},
\end{align}
where $f^{'}(r_0)=0$ and $f(r) \approx f(r_0)+\frac{1}{2}(r-r_0)^2 f^{''}(r_0)$. Recognizing $f(r) = i \int_0^{r}d \tilde{r} \frac{m_{A'}^2 - \omega_p^2(\tilde{r})}{2k} $,
the probability $P_{A'\to \gamma}$ in Eq.~\eqref{eq:Pfinal2} can be simplified to Eq.~\eqref{eq:Pfinal}. One can explicitly expand $f(r)$ to the next order and show that the correction is about $f'''(r_0)/(f''(r_0))^{3/2}\approx v(r_c)/(k \Delta r_c)^{1/2}$, where $v(r_c)$ is the dark photon velocity at the resonant region, $k^{-1}$ can be seen as the de Broglie wave length, and $\Delta r_c$ is the resonant length. 
The dark photon velocity  is about $10^{-3}$ times the speed of light.
The de Broglie wavelength of the dark photon is about $0.1-10$ km.
The size of the resonant region is at the scale of about $10^3$ km.   
Therefore, the next-leading order effect in our case is suppressed by a factor of $10^{-5}$.

Clearly, the wave method is in good agreement with
the quantum field method, which calculates only the resonant contribution.
Another way to understand this is that outside the resonant region, the phase $e^{-i\int d\tilde{r}\cdots}$ in 
Eq.~\eqref{eq:Pfinal2} oscillates quite fast, which cancels themselves in the probability amplitude.

Besides this linearized equation technique, one may also solve it similarly as neutrino oscillations 
with the mass matrix given in Eq.~\eqref{eq:wave1}, see Ref.~\cite{Mirizzi:2009iz}.
The result is in agreement with the above two methods . 

\bigskip

\noindent \textbf{The solar model--}
The $A'\rightarrow \gamma$ conversion happens in the solar corona. Like the atmosphere 
on  the Earth, it is a complex and vibrant environment. The corona can be divided into three regions.
The active region holds most of the activities but makes up only a small fraction of the total surface area, like the cities on the Earth. The coronal hole region are the northern and southern polar zones of the Sun. The quiet Sun region is the rest of the surface area, which is not static but has minor dynamic processes with small scale phenomena comparing to the active regions. 

We focus on the quiet Sun region for our study, because it has less active events like solar flares. Although it is not 
fully quiet, with some minor dynamic processes, we model it as a spherical symmetric and hydrostatic, 
in which the gas pressure is balanced by the gravitational force and is static in time. 
Indeed, the quiet Sun region does show perfect hydrostatic equilibrium, see Refs.~\cite{1981ApJS...45..635V, Aschwanden_2001}.

The relevant quantities in our calculations are the electron number density $n_e$ and temperature $T$ profiles. 
We take the profiles from Ref.~\cite{2008GeofI..47..197D}, where they have calculated the temperature $T$ and hydrogen density $n_H$ profiles for the quiet sun regime based on photospheric model from Ref.~\cite{1981ApJS...45..635V} and coronal model from Ref.~\cite{1976RSPTA.281..339G, peter1990solar}. 
With spherical symmetry assumption, hydrostatic equilibrium and radiative transfer assumption, they calculated 
the electron number density profile $n_e$. We have not used the Pakal code developed in Ref.~\cite{2008GeofI..47..197D}, but only $n_e$ and $T$ profiles which are the input for Pakal code.
Those profiles have also been calculated by different groups~\cite{aschwanden2006physics}
using chromosphere model from Ref.~\cite{1990ApJ...355..700F} and again coronal model from Ref.~\cite{1976RSPTA.281..339G, peter1990solar}. 
Their results are in agreement with each other.

More importantly, their predictions on profiles have been verified by various atomic 
lines observations at soft X-ray range~\cite{Aschwanden_2001} and extreme ultra-violet range \cite{1981ApJS...45..635V}. For example, the $n_e$ profile for quiet Sun is  in good agreement with the various observations~\cite{Aschwanden_2001} 
and the $T$ profile gives the temperature in the right
range (1--2 million Kelvin)~\cite{Aschwanden_2001} comparing with the extreme-ultraviolet line observations \cite{1996ApJS..106..143B}.
Therefore, the profiles used in the paper are simple and reliable.

The spherical and hydrostatic profile or model we used for solar atmosphere is not the most recent one,
but is simple and consistent with the atomic line observations. The more recent development of the solar atmospheric model
includes changing from hydrostatic equilibrium to hydrodynamic, by adding the continuity equation due to particle number conservation. 
The magnetic field is also very important for plasma movement. The combination of these effects is called magneto-hydrodynamics (MHD) model.
For example, the particles not energetic enough will flow along the magnetic flux line.
So, the plasma is treated as a fluid governed by gravitational, electromagnetic interactions.
However, since we are only focusing on the quiet Sun region, and the relevant quantities are the density and temperature profiles, 
we believe that the spherical and hydrostatic model already provides a good description of the quiet Sun region.
\\

\noindent \textbf{The uncertainties in the calculation--}
Regarding the uncertainties from the solar model, the relevant errors come from $n_e$ and $T$ profiles.
The $n_e$ profile for quiet Sun is good within a factor of a few from the various observations~\cite{Aschwanden_2001}.
Its square root determines the plasma frequency, which only shifts the location of resonant region. 
Its derivative on radius determines the
conversion probability and the slope does fit nicely with the observational data~\cite{Aschwanden_2001}.

On the other hand, the $T$ profile determines the absorption of photon from inverse bremsstrahlung process,
where the absorption rate is proportional to $T^{-3/2}$. 
The column emission measure, which is proportional to the line-of-sight integral of $n_e^2$, can be extracted from the 
broad range of extreme-ultraviolet and soft X-ray line observations. Its differential distribution over temperature 
for the model prediction~\cite{Aschwanden_2001} and the extreme-ultraviolet observation data~\cite{1996ApJS..106..143B} are peaked around $\log_{10} ({\rm T [Kelvin]}) \sim 6.3 $ and
$6.1-6.2$ respectively, for low corona of quiet Sun. Therefore, the temperature profile is in pretty 
good agreement with data. 

The other uncertainty in solar model is related to the spherical symmetric and hydrostatic assumption.
In reality, the Sun has a vibrant environment, that the turbulences and flares in the corona can make 
$n_e$ non-spherical and even evolve with time. This will distort the spherical distribution of $n_e$
and leads to non-radial photon propagation direction outside the Sun.
There are several reasons to alleviate the above concerns. Firstly, we have already chosen the quiet Sun region, which
has the least dynamic activities comparing to the active regions. Second, the hydrostatic assumption together
with spherical symmetry has been explicitly tested by Refs.~~\cite{1981ApJS...45..635V, Aschwanden_2001}
using soft X-ray and extreme-ultraviolet line observations. Therefore, the static and spherical symmetric picture 
is a good approximation in the sense of time and spatial average for the quiet Sun region. 
Thirdly, the activities can modify the above quantities but should not have preferred directions, 
unless there are underline substructures. Therefore, in the sense of the spatial average,
out-going direction of the converted photon is isotropic.

As a result, we summarize for the solar corona model that we have used a simple model for the quiet sun. It is
hydrostatic and spherical symmetric, but it is sufficient for our purpose of DM search. It only needs 1D profile which significantly simplify the signal calculation and the uncertainty for this model should be within a factor of a few.

Next, we move to the uncertainties from  dark matter model. The first uncertainty is the local DM density. We
have used the value $\rho_{\rm DM} = 0.4 ~{\rm GeV}{\rm cm}^{-3}$, which is an average number from the N-body
simulation from the DM study. It is possible that the solar system sits in the DM substructure, that the density
is boosted than other region. It is also possible that the density is much smaller than the average value due to fluctuation
from the structure formation. As a result, the density provides the uncertainty as large as a factor of few.

The second possible uncertainty is from the local DM velocity. In our calculation, the inverse of velocity $v^{-1}$ from
conversion probability is canceled by the velocity in the DM flux, when calculating the radiation power. 
Therefore, the signal is less affected by the DM velocity comparing with the density.

Therefore, we conclude that the uncertainties from  DM model that provides uncertainties from its density, which is
a factor of a few. Together with the uncertainties from  solar model, the predicted signal has an uncertainty of
a few and should be within one order.

\end{document}